\documentstyle{article}

\begin{document}

\renewcommand{\baselinestretch}{1.5}
\large


\title{GUTs and String-GUTs}
\author{Abdel P\'erez--Lorenzana$^{1,2}$ and William A. Ponce$^{3}$}

\date{\today}

\maketitle

\vspace*{-15em}\begin{flushright}{UMD-PP-99-115}\end{flushright}
\vskip10em

\begin{center}
\normalsize 1-Department of Physics, University of Maryland, College
\normalsize Park, MD 20742 USA. \\
\normalsize 2-Departamento de F\'\i sica, 
\normalsize Centro de Investigaci\'on y de Estudios Avanzados del I.P.N.\\
\normalsize Apdo. Post. 14-740, 07000, M\'exico, D.F., M\'exico.\\
\normalsize 3-Departamento de F\'\i sica, Universidad de Antioquia 
\normalsize A.A. 1226, Medell\'\i n, Colombia.
\end{center}

\begin{abstract}
We look for a connection between string theories and Grand Unified
Theories (GUTs), with the aim to look for new insights in the existing
four dimensional string-GUT problems. We argue that the construction of
consistent string-GUT models could require the use of non canonical affine
levels.  We list the most common level values related to realistic GUTs.

 \end{abstract}

Pacs: {11.25.Mj;12.10.-g;12.10.Kt}

\vskip2em

\section{Strings}
Strings provide us with a very compelling theory, giving a consistent
framework which is finite and incorporates at the same time both, quantum
gravity and chiral supersymmetric (SUSY) gauge theories. When one-loop
effects are included in the perturbative heterotic string\cite{gin} 
they predict a unification of the
gauge couplings at a scale $M_{string}\sim 4\; \times 10^{17}$ GeV. 

Unification of coupling constants is a necessary phenomenon in string
theory. Specifically, at tree level, the gauge couplings 
$\alpha_i=g_i^2/4\pi$,
($i=1,2,3$, for the groups of the Standard Model (SM) 
factors $U(1)_Y,\;\; SU(2)_L$, and
$SU(3)_c$ respectively)  are related at the string scale by\cite{ginn}
\begin{equation}
\kappa_3\alpha_3=\kappa_2\alpha_2=\kappa_1\alpha_1,   \label{kac}
\end{equation}
where $\kappa_i, \; i=1,2,3 $ are the affine levels, or Kac-Moody
levels, at which the group factor $U(1)_Y,\; SU(2)_L$, and $SU(3)_c$ is
realized in the four-dimensional string.

To calculate the Kac-Moody levels, 
the starting point is the ten-dimensional heterotic string with gauge
group $SO(32)$ or $E_8\otimes E_8$ corresponding to an affine Lie algebra
at level $\kappa = 1$. A standard compactification\cite{orbi} leads to a
four dimensional model with gauge group formed by a product of non-abelian
gauge groups $G_i$ realized at levels $\kappa_i =1$, times $U(1)$ factors.
Building string theories with non-abelian algebras at higher levels
($\kappa = 2,3,\dots$) is considerable more difficult than at level one,
and new methods for compactification must be developed\cite{alda} (to
produce levels beyond $\kappa = 3$ is a very cumbersome task). Now, the
affine levels for abelian $U(1)$ factors can not be determined from
algebraic procedures and their values may be considered as free parameters
in the four dimensional string\cite{iba}. 

Then, the compactification of the heterotic string to the four dimensional
$G_{SM}\equiv SU(3)_c\otimes SU(2)_L\otimes U(1)_Y$ 
could be achieved at $M_{string}$, with
$\kappa_2,\kappa_3=1,2,\dots n$, an integer number, and $\kappa_1$ a
normalization free coefficient ($\kappa_1 >1$ in order for
the $e_R$ to be in the massless spectrum of the four
dimensional string\cite{schelle}). The compactification to a four
dimensional simple gauge group $G(=SU(5),\; SO(10),\; E_6$, etc.) has also
been partially studied in the literature, with upper values for the
integer $\kappa$ levels calculated\cite{ellis}. Also, strings with
$SU(5)\subset SU(5)\otimes SU(5)$ and $SO(10)\subset SO(10)\otimes SO(10)$
at levels $\kappa_2=\kappa_3=2$ have been presented in Ref.\cite{alda}.

The  values attained by level $\kappa_i$ play a fundamental role
in  string theories, because they fix at the string scale the electroweak
mixing angle $\sin \theta_W$. Besides, they impose limits on possible
representations allowed at low energies\cite{alda}, and determine the
conformal spin of the currents $J$ which are forced to be in the spectrum
because of charge quantization\cite{schelle}. So, theories with different
$\kappa_i$ values must have quite different physical implications.

Today it is believed that $M_{string}$ could be not  the perturbative
value $4\times 10^{17}$ GeV, but a smaller one (maybe as small as 1
TeV)~\cite{qui} coming from the non perturbative effects of the string.
This matter has not been settled yet, and it is not crucial for the
analysis which follows.

\section{GUTs}

In a particular GUT model, the unification of the three SM gauge
couplings is properly achieved if they meet together into a common value
$\alpha=g^2/4\pi$ at a certain energy scale $M$, where $g$ is the gauge
coupling constant of the unifying group $G$. However, since $G\supset
G_{SM}$, the normalization of
the generators corresponding to the subgroups $U(1)_Y,\; SU(2)_L$, and
$SU(3)_c$ is in general different for each particular group $G$, and
therefore the SM coupling constants $\alpha_i$ differ at the unification 
scale from $\alpha$ by numerical factors $c_i\;(\alpha_i=c_i\alpha)$. As a
matter of fact, if $\alpha_i$ is the coupling constant of $G_i$, a simple
group embedded into $G$, then 
\begin{equation}
c_i\equiv\frac{\alpha_i}{\alpha}=\frac{{\rm Tr.}T^2}{{\rm Tr.}T_i^2}
\label{traces}
\end{equation}
where $T$ is a generator of the subgroup $G_i$ properly normalized over a
representation $R$ of $G$, and $T_i$ is the same generator but normalized
over the representations of $G_i$ embedded into $R$ (the traces run over
complete representations); so, if just one standard
doublet of $SU(2)_L$ is contained in the fundamental representation of $G$
(plus any number of $SU(2)_L$ singlets), then $c_2=1$ (as in
$SU(5)$ \cite{su5} for example). In this way we proof that for $i=2,3, \;
c_i^{-1}$ is an integer number. 

The constants $c_i$ are thus pure rational numbers satisfying $c_1>0$, and
$0<c_{2(3)}\leq 1$. They are fixed once we fix the unifying gauge
structure, 
and from pure algebraic arguments we must have at the GUT scale
\begin{equation}
c^{-1}_3\alpha_3=c^{-1}_2\alpha_2=c^{-1}_1\alpha_1 .  \label{gut}
\end{equation}

In Table 1 we present  the $c_i$ $i = 1; 2; 3$ values  for most of the GUT
groups in the literature; they are calculated using Eq.(\ref{traces}). The
canonical entry is associated with the following nine groups: 
$SU(5)$\cite{su5}, $SO(10)$\cite{so10}, $E_6$\cite{e6}, $[SU(3)]^3\times
Z_3$\cite{3su3}, $SU(15)$\cite{su15}, $SU(16)$\cite{su16}, $SU(8)\times
SU(8)$\cite{2su8},  $E_8$\cite{e8}, and $SO(18)$\cite{so18}. The model
[SU(3)]$^4\times Z_4$ is taken from Reference\cite{4su3}, 
$SU(5)\otimes SU(5)$ from\cite{2su5}, $SO(10)\otimes SO(10)$
from\cite{2so10}, $[SU(6)]^3\times Z_3$ from\cite{3su6}, $[SU(6)]^4\times
Z_4$ from\cite{4su6}, $E_7$ from\cite{e7}, $[SU(4)]^3\times Z_3$
from\cite{2so10}, and $[SU(2F)]^4\times Z_4$ (the
Pati-Salam models for $F$ families) from\cite{patis}.

In the canonical entry we have normalized the $c_i$ values for some
groups  to the SU(5) numbers; for example, the actual values for SO(10)
are $\{c^{-1}_1; c^{-1}_2; c^{-1}_3\} = \{{10/ 3};2; 2\} =2\{5/3; 1;
1\}$, and for SU(16) are 
$\{c^{-1}_1; c^{-1}_2; c^{-1}_3\} = \{{20/ 3};4; 4\} =4\{5/3; 1; 1\}$. 
This normalization makes sense because the common factor can be absorbed
in the GUT coupling constant $\alpha$; besides, physical
quantities such as $\sin^2\theta_W$, $M_{GUT}$, etc., depend only on
ratios of the $c_i$ values.

$c^{-1}_3$ can take only the values $1,2,3,4$ for one family groups,  or
higher integer values for family groups. $c^{-1}_3 = 1$ when it is
$SU(3)_c$ which is embedded in the GUT group $G$; $c^{-1}_3 = 2$ when it
is the chiral color~\cite{2su3} $SU(3)_{cL}\times SU(3)_{cR}$ which is
embedded in $G$, etc. For example $c^{-1}_3 = 4$ in SU(16) due to the fact
that the color group in SU(16) is $SU(3)_{cuR}\times SU(3)_{cdR}\times
SU(3)_{cuL}\times SU(3)_{cdL}$.

For some family groups $c^{-1}_2$ take the values $1, 2,\dots F$ for $1,
2, \dots F$ families.  Indeed, the $c_i$ values for the $F$ family
Pati-Salam models~\cite{patis} $[SU(2F)]^4\times Z_4$ are 
$\{c^{-1}_1; c^{-1}_2; c^{-1}_3\} =  \{(9F - 8)/3; F; 2\}$; 
and for $[SU(2F)]^3\times Z_3=SU(2F)_L\otimes SU(2F)_c\otimes
SU(2F)_R\times Z_3$ (the $2F$ color vectorlike version of the Pati-Salam
models\cite{www}), $\{c^{-1}_1; c^{-1}_2; c^{-1}_3\} =  \{(6F - 4)/3; F;
1\}$. 

In general, $c^{-1}_{2(3)}=1,2,\dots f$, where $f$ is the number of
fundamental representations of $SU(2)_L$ $(SU(3)_c)$ contained in the
fundamental representation of the GUT group. For example, $c_2^{-1}=4$ in
$SU(16)$ because the 16 representation of $SU(16)$ contains four $SU(2)_L$
doublets; three for $(u,d)_L$ and one for $(\nu_e,e)_L$. 

The group $[SU(4)]^3\times Z_3$ in Table 1 is not the vector-like color
version of  the two family Pati-Salam model,  but it is the one family
model introduced in Ref.~\cite{2so10}. The  group $[SU(6)]^4\times Z_4$ in
the Table could be the three family Pati-Salam model\cite{patis}, or
either the version of such a model without mirror fermions introduced in
Ref.~\cite{4su6}. All models in Table 1 are realistic, except
$E_7$\cite{e7} which is a two family model with the right handed quarks in
$SU(2)_L$ doublets.

Notice that the values for $c^{-1}_1$ are integer multiple of 1/3 for all
the groups in the table, which is due to the condition for having only
standard electric charges in the representations of the particular group
used as a GUT. Such condition reads
\begin{equation}
c_1^{-1} + c_2^{-1} + {4\over 3} c_3^{-1} = 0 \quad mod.\ 4
\end{equation}
which is satisfied by all entries in the table (in some entries the real
values must be used instead of the normalized ones).

\section{String-GUTs}

The logarithmic running through the ``desert'' of the fundamental 
coupling constant
is governed by the following renormalization group equations:
\begin{equation}
\alpha_i^{-1}(\mu)=\eta_i\alpha^{-1}-{b_i\over 2\pi}\ln\left({M\over
\mu}\right)+\Delta_i
\label{rge}
\end{equation}
where $b_i$ are the one-loop 
beta functions, M the unification scale and $\Delta_i$
the threshold and other corrections.

GUTs (and SUSY-GUTs)  were invented\cite{su5} before strings, 
and they may exist by
themselves as independent physical entities. For the several GUT models
$\eta_i=c_i^{-1}$ in Eq. (\ref{rge}), $M=M_{GUT}$ is the GUT 
scale, and $\alpha
= g^2/4\pi$, with $g$ the coupling constant of the GUT group. However, 
it is a well known result that 
 the logarithmic  running through the desert of the three gauge
couplings $c_i\alpha_i^{-1}$ for the canonical values $\{c_1,c_2,c_3\} = 
\{{3\over 5},1,1\}$,  do merge together into a single point, only when the
SUSY partners of the SM elementary particles are included in the
renormalization group equations at a mass scale $M_{susy}\sim
1$TeV\cite{susy}.  This amazing result  is not upset when higher order
contributions are taken into account\cite{hig},  and it provides the
unification scale $M_{GUT}\sim 2\; \times 10^{16}$GeV.

Now, strings may exist without GUTs. If this is the case, then the string
must compactify to four dimensions to the SM gauge structure $G_{SM}$  at
a mass scale $M_{string}$. The fundamental coupling constants still run
according to Eq. (\ref{rge}), where now $M=M_{string}$, $\eta_i=\kappa_i$
and
$\alpha=g^2_{string}/4\pi$.
When we solve Eqs.(\ref{rge}) using the one loop SUSY beta functions
$(b_1,b_2,b_3) = (-11,-1,3)$, and  the canonical Kac-Moody levels
$\{\kappa_1,\kappa_2,\kappa_3\} = \{ {5\over 3},1,1\}$,
we get
$M_{string}= 2\times 10^{16}$GeV,
 a factor of 20 smaller than its perturbative value. If we solve the
equation for $M=M_{string}= 4\times 10^{17}$GeV, we get  $\kappa_1 =
{5\over 4}$\cite{wp}. This is the so call string-GUT problem\cite{dien}
(the inclusion of the second order corrections does not seem to solve the
problem either\cite{ibann}).

But it may happen that string-GUTs are real objects. They exist if the
string compactifies in four dimensions not to $G_{SM}$, but to a simple
group  $G$ that acts as a unified group. If this is the case, then $M=
M_{GUT}=M_{string}$, and $\eta_i=c_i^{-1}=\kappa_i$, which become two
necessary conditions for having a consistent string-GUT. In this kind of
theories not only the entire unification of interactions is realized, but
also the advantages of the GUT symmetry are available,  as for example 
possible explanation of  fermion masses and mixings angles,  neutrino
oscillations, and other features not provided by the string alone.

\section{Final Remarks}

In this note most of the four dimensional string Kac-Moody levels which
could be related to GUT theories are presented in Table 1 (a guide for
string-GUT model builders). From the table we may visualize the wide
spectrum available for the values $\kappa_i = c_i^{-1}, \; i=1,2,3$. 

So far, almost the entire literature on four dimensional strings has been
focused on the canonical values $\kappa_2=\kappa_3=1,\;\kappa_1=5/3$,
pointing towards a canonical string-GUT model, or to a string model
without a relation to a particular GUT. But  as it is known, there are
serious problems with the models constructed so far. Just to mention a few
we have: the string-GUT problem\cite{dien}, the doublet-triplet
problem\cite{alda}, the failure to produce a consistent low energy
particle spectrum\cite{alda}, etc. It may be feasible that the
construction of four dimensional string theories with non-canonical
$\kappa_i$ values may ameliorate, or even cure some of the mentioned
problems (in the model of Ref.\cite{4su3}, $M_{GUT}\geq M_{string}$, and
the doublet triplet problem is not present at tree level).

$\kappa_i,\; i=1,2,3$ values, different from the canonical ones, are in
general related to the existence of non standard matter. That extra matter
can have a mass at an intermediate scale, or either at the string-GUT
scale (most of the string-GUT models constructed so far contain non
standard matter).

{\bf Concluding remark}: look for string-GUTs with $\kappa_i$  levels
other than the canonical ones.

\section*{Acknowledgments}
This work was partially supported by CONACyT, M\'exico and Colciencias,
Colombia.  W.A.P. thanks  R.N. Mohapatra, and the Physics Department of
the University of Maryland at College Park for hospitality during the
completion of this work.

\begin{table}
\begin{center}
\begin{tabular}{||l|c|c|c||}
\hline
Group & $c^{-1}_1$  & $c^{-1}_2$ & $c^{-1}_3$\\  \hline\hline
{\bf CANONICAL} (9 groups)  &5/3   &1  &1 \\ \hline
[SU(3)]$^4\times Z_4$ & 5/3 & 1 & 2 \\ \hline
SU(5)$\otimes$ SU(5), SO(10)$\otimes$SO(10) & 13/3  & 1  & 2 \\ \hline
$[SU(6)]^3\times Z_3$ & 14/3 & 3 & 1 \\  \hline
$[SU(6)]^4\times Z_4$ & 19/3 & 3 & 2 \\ 
\hline
$E_7$  & 2/3 & 2 &  1 \\
\hline
$[SU(4)]^3\times Z_3$ & 11/3 & 1 & 1 \\
\hline
$[SU(2F)]^4\times Z_4$ & $(9F-8)/3$ & $F$ &  2 \\ \hline
$[SU(2F)]^3\times Z_4$ & $(6F-4)/3$ & $F$ &  1 \\
\hline
\end{tabular}
\caption{
{$c_1, c_2$ and $c_3$ 
values for most of the GUT models in the literature. $F=1,2,\dots$ stands
for the number of families in that particular model. The 9 ``canonical"
groups are presented in the main text.}}
\end{center}
\end{table}


\end{document}